\documentclass[preprint,12pt]{elsarticle}
\usepackage{amsmath}
\usepackage{amssymb}
\usepackage{graphicx}

\journal{Physica D}

\begin{document}

\begin{frontmatter}

\title{The synchronization transition in correlated oscillator populations}

\author{Markus Brede}
%\thanks{}
\address{CSIRO Marine and Atmospheric Research, CSIRO Centre for Complex System Science, F C Pye Laboratory,
GPO Box 1666, Clunies Ross Street
Canberra ACT 2601, Australia}

%\email{Markus.Brede@Csiro.au}


\date{\today}

\begin{abstract}
The synchronization transition of correlated ensembles of coupled Kuramoto oscillators on sparse random networks is investigated. Extensive numerical simulations show that correlations between the native frequencies of adjacent oscillators on the network systematically shift the critical point as well as the critical exponents characterizing the transition. Negative correlations imply an onset of synchronization for smaller coupling, whereas positive correlations shift the critical coupling towards larger interaction strengths. For negatively correlated oscillators the transition still exhibits critical behaviour similar to the all-to-all coupled Kuramoto system, while positive correlations change the universality class of the transition depending on the correlation strength. Crucially, the paper demonstrates that the synchronization behaviour is not only determined by the coupling architecture, but is also strongly influenced by the oscillator placement on the coupling network.
\end{abstract}

%\pacs{05.45.Xt,89.75.Fb,89.75.-k}
%\keywords{Complex Networks;Synchronization;Kuramoto Model}

\end{frontmatter}

\maketitle

\section{Introduction}

The collective dynamics of synchronization has constituted a field of very active interest over the last couple of years. Typical areas where synchronization phenomena play an important role are manifold, ranging from fields as diverse as ecology, social dynamics, biological rhythms to fields as laser physics \cite{Winfree,Kuramoto,Manrubia} and also power systems \cite{DH}. These real-world systems consist of many elementary units that interact. Since the discovery that the coupling networks are often non-trivial \cite{nets,Boc0}, the research about synchronization phenomena on complex networks has attracted significant attention \cite{Boc0,Ar0}.

So far most of the work in this field has dealt with understanding the influence of the topology of the interaction network on synchronization properties \cite{Boc0,Ar0}. Studies of this problem have mainly concentrated on three approaches: (i) the master stability function approach \cite{Pecora} to understand the stability of the fully synchronized state for systems of identical chaotic oscillators; (ii) several analytical methods to study the onset of synchronization \cite{Restrepo,Lee}, and; (iii) a numerical exploration of the properties of the synchronization transition of network ensembles \cite{Hong,Hong1,Hong2,HongL,Moreno,Gardenez,Arenas,Me}. These studies have provided much insight, demonstating, e.g., that for symmetrical coupling small homogeneous uncliquish load-balanced networks facilitate the transition to complete synchronization \cite{Donetti}. In asymmetrically coupled systems, which have mainly been investigated through weighting schemes on undirected networks \cite{Motter,Zhou,Hwang,Chavez}, homogeneous in-signals and balanced loads constitute two key indicators of an enhanced synchronizability.

In the second and third stream of this research the critical coupling that characterizes the onset of synchronization and its relation to structural properties of the coupling network have been explored, mainly using the Kuramoto model \cite{Kuramoto}
\begin{align}
 \label{Kuram}
 \dot{\phi_i}=\omega_i+\sigma \sum_{j} a_{ij} \sin(\phi_j-\phi_i)
\end{align}
 as a well-understood model for the study of phase synchronization (see, e.g., \cite{Acebron} for a recent summary). In Eq. (\ref{Kuram}) the $\phi_i$, $i=1...N$ describe the phases of $N$ oscillators, the $\omega_i$ their native frequencies, $\sigma$ the coupling strength and the matrix $a_{ij}$ the topology of the coupling, i.e. the interaction network. Following, e.g., \cite{Restrepo,Moreno,Ichinomiya,Buzna} we set $a_{ij}=1$ if $i$ and $j$ are connected and $a_{ij}=0$ otherwise. For a very good discussion of this choice of normalization see \cite{Arenas}.

 For all-to-all coupling $a_{i,j}=1/(N-1) \forall i\not=j$ the model (\ref{Kuram}) exhibits a second order phase transition from a desynchronized to a (partially) synchronized phase at some critical coupling strength $\sigma_c=2/\pi g(0)$ \cite{Kuramoto}, where $g(\cdot)$ is the distribution of the oscillator's native frequencies. Interestingly, the synchronization transition appears to be very similar to the  mean-field type version of the model for some classes of complex networks such as the Strogatz-Watts small world model \cite{Hong}, random graphs, and even some types of scale-free networks \cite{Moreno}. However, for some types of degree distributions of scale-free networks the characteristics of the synchronization transition are found to depend on the degree heterogeneity \cite{Lee}.

Recently, as another approach, optimization techniques were used to investigate the relationship between correlations in oscillator placements, network architecture and synchronization \cite{MB1,MB2,MB3,Carareto}. The results from these studies suggest that a correlated oscillator placement can strongly affect a system's synchronization properties. More specifically, a grouping of oscillators on the network in such a way that the native frequencies of linked oscillators are anti-correlated has been shown to induce a transition to complete synchronization for low coupling, whereas a positively correlated oscillator arrangement requires more coupling strength for macroscopic synchronization to occur. These findings have recently been corroborated by a study of synchronization in bipolar population networks of Kuramoto oscillators \cite{Buzna}.

The results in \cite{MB1,Carareto,Buzna} suggest that the characteristics of the transition to synchronization may be different for positively or negatively correlated oscillator arrangements. However, this issue and the scaling with the system size has not been studied systematically so far and is the main subject of the present study.

In our view, these results appear particularly relevant for evolved ecological or biological systems, where the synchronization of the individual elements might have been a determinant of the fitness of the system and thus have guided its evolution. In such a case, the evolution of the system will probably have been strongly affected by heterogeneity in the characteristics (i.e. native frequencies in the Kuramoto model) of the individual elements. Thus, the observed architecture of the system is not only characterized by the topology of the interaction network, but also strongly determined by the oscillator placement on the network. Similarly, as \cite{MB1} highlights that gains in the average degree of synchronization are at least as strongly marked by rearrangements in the oscillator placement as by the evolution of the network topology itself, introducing specific oscillator correlations may be easier to implement than a change in the overall network arrangement in technical applications where synchronization between the individual elements is desirable.

Correlations between the native frequencies of adjacent oscillators can be measured by a standard Pearson correlation coefficient
\begin{align}
\label{CW}
 c_\omega = \frac{\sum_{i,j} a_{ij}(\omega_i-\langle \omega \rangle)(\omega_j-\langle \omega \rangle) } {\sum_{i,j} (\omega_i-\langle \omega \rangle)^2 a_{ij}},
\end{align}
where $\langle \omega\rangle=1/N\sum_i \omega_i$ is the average native frequency. It should be noted that strongly correlated oscillator placements are only possible on sparsely connected interaction networks. Generally, the more connected a network is, the less correlated oscillator placements are possible, i.e. for a fully coupled system one trivially has $c_\omega=0$. To our best knowledge, apart from \cite{MB1,MB2,MB3}, correlations between network topology and oscillator placement and their influence on synchronization have not found any attention in the considerable literature on synchronization of complex networks, cf., e.g. \cite{Boc0,Ar0}, for a recent survey.

In this paper, we follow up on the above findings of our previous work  and carry out a detailed analysis of the properties of the synchronization transition on random networks with tuneable correlated oscillator placements. As will be shown below, correlations in the oscillator placement can affect both the critical point that separates the desynchronized from the synchronized phase, as well as the universality class of the synchronization transition. Results that elaborate the dependence of the properties of the transition on the characteristics of the oscillator arrangement on the network are detailed below.

\section{Analysis of the synchronization transition}

To investigate the synchronization transition we follow a straightforward approach and construct random (undirected) Erd\"os-R\'enyi type graphs \cite{ER}. For technical reasons, to eliminate an irrelevant source of heterogeneity that otherwise would have to be averaged over, we choose the `microcanonical' random graph model, where exactly $L=pN$ distinct randomly chosen pairs of nodes are connected by  links \cite{remark}. Starting from an initially random oscillator placement we then carry out a simple optimization procedure to generate a correlated oscillator arrangement with a given correlation coefficient $c_\omega^*$. This is achieved by randomly selecting pairs of nodes and suggesting to swap the associated native frequencies. Swaps are accepted if they lead to a correlation coefficient $c_\omega$ closer to the desired value $c_\omega^*$, i.e. we minimize the difference $|c_\omega-c_\omega^*|$. The procedure is terminated if the correlation coefficient reaches the desired level of correlation within a small error tolerance, i.e. $|c_\omega-c_\omega^*|\ll \epsilon$, where we set $\epsilon=10^{-4}$ for the rest of the study. Once the correlated oscillator ensemble is generated, we then integrate Eq.'s (\ref{Kuram}) numerically with initial conditions $\phi_i(t=0)$ randomly selected from $(-\pi,\pi]$ and determine the standard order parameter
\begin{align}
r(t) e^{i\psi}=\sum_j e^{i\phi_j(t)}.
\end{align}
The amplitude order parameter $r(t)$ is then averaged over time (after allowing for a sufficient time interval for relaxation) and then over several hundred different (correlated) oscillator $\omega_i$ and network configurations $a_{ij}$, where the native frequencies $\omega_i$ are selected uniformly at random from $[-1,1]$ \cite{Bem0}. More precisely, we measure
\begin{align}
 r=\langle 1/T\sum_{t=T_\text{rel}}^{T_\text{rel}+T} r(t)\rangle,
\end{align}
where $\langle \cdot \rangle$ indicates the average over different initial conditions and the oscillator and network ensemble. As, e.g., in \cite{Hong,Hong1,Hong2,Gardenez,Me} a finite size scaling analysis is then carried out to determine the characteristics of the synchronization transition. 

In the following analysis we focus on networks with average degree $\langle k\rangle=3.5$, i.e. very sparsely connected networks which nevertheless already have a giant component that comprises more than $95\%$ of all nodes. Clearly, oscillators cannot be placed on loops of odd length in a perfectly anti-correlated manner.  Thus, large densities of short loops of odd length don't allow for strongly anti-correlated oscillator placements on the network.  In this context, the sparse connectivity ensures an asymptotically vanishing number of triangles and in general low densities of short loops. It is worthwhile to point out that in order to highlight the effect of oscillator correlations on the synchronization transition in a regime in which the effect is strongest, all experiments have been carried out on very sparsely connected networks. In the light of this, comparisons to previous numerical and analytical work via mean-field approaches \cite{Hong1,Hong2,HongL} are difficult.

Using the above optimization method correlated oscillator placements with $c_\omega$ in the range between $-0.6$ and $0.6$ can be generated. Even though the rest of the study is focussed on only one value of network connectivity $\langle k\rangle=3.5$, we have also experimented with different link densities in this sparse regime. The results presented below are found to be prototypical and the qualitative statement appears to be generally valid for sparsely connected networks.

\begin{figure}
 \begin{center}
 \includegraphics[width=.7\textwidth]{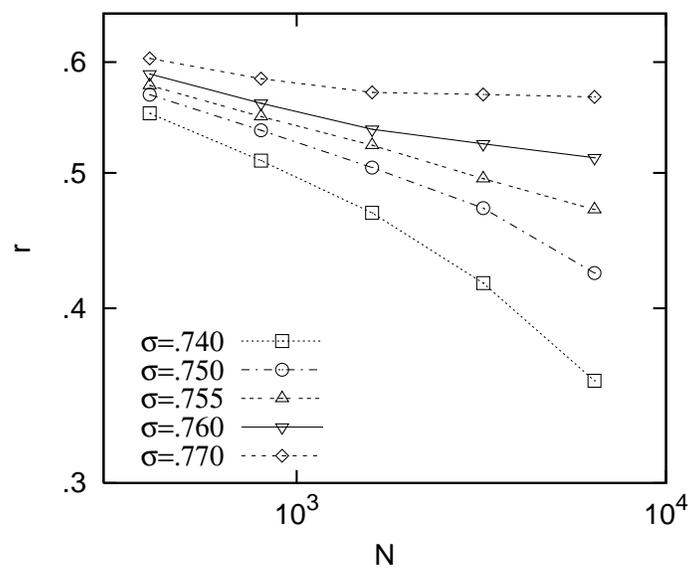}
 \caption{Scaling of the order parameter with the system size for various coupling strengths for $c_\omega=0.6$. For $\sigma\approx 0.755$ the decay becomes algebraic.}
\label{fig.1a}
 \end{center}
 \end{figure}

The precise determination of the transition between the desynchronized and the synchronized phase requires a careful consideration of finite size effects. Here we follow the approach of \cite{Hong,Gardenez,Me} with the finite size scaling ansatz
\begin{align}
\label{FSS}
 r(\sigma,N)=N^{-\alpha} F((\sigma-\sigma_\text{c})N^{1/\nu}),
\end{align}
where the exponent $\alpha=\beta/\nu$  is related to the exponent $\beta$ that describes the scaling of the order parameter $r$ close to the critical point in the thermodynamic limit, i.e.
\begin{align}
 r \sim (\sigma-\sigma_\text{c})^\beta.
\end{align}
The exponent $\nu$ in Eq. (\ref{FSS}) characterizes the divergence of the correlation volume at the critical point and the scaling function $F (x)$ is bounded in the limit $x\to \pm\infty$.

From the ansatz (\ref{FSS}) the critical point $\sigma_\text{c}$ can be determined by plots of $rN^{\alpha}$ vs. $\sigma$ for varying system sizes and different values of the exponent $\alpha$. For the correct choice of the exponent $\alpha$ the curves $Nr^\alpha (\sigma)$ intersect at a unique crossing point. Having thus determined the critical coupling and the exponent $\alpha$, the exponent $1/\nu$ can be obtained as the exponent that allows for a collapse  of data points for different system sizes when plotting $rN^\alpha$ vs. $(\sigma-\sigma_c)N^{1/\nu}$.

\begin{figure*}
 \begin{center}
\includegraphics[width=.95\textwidth]{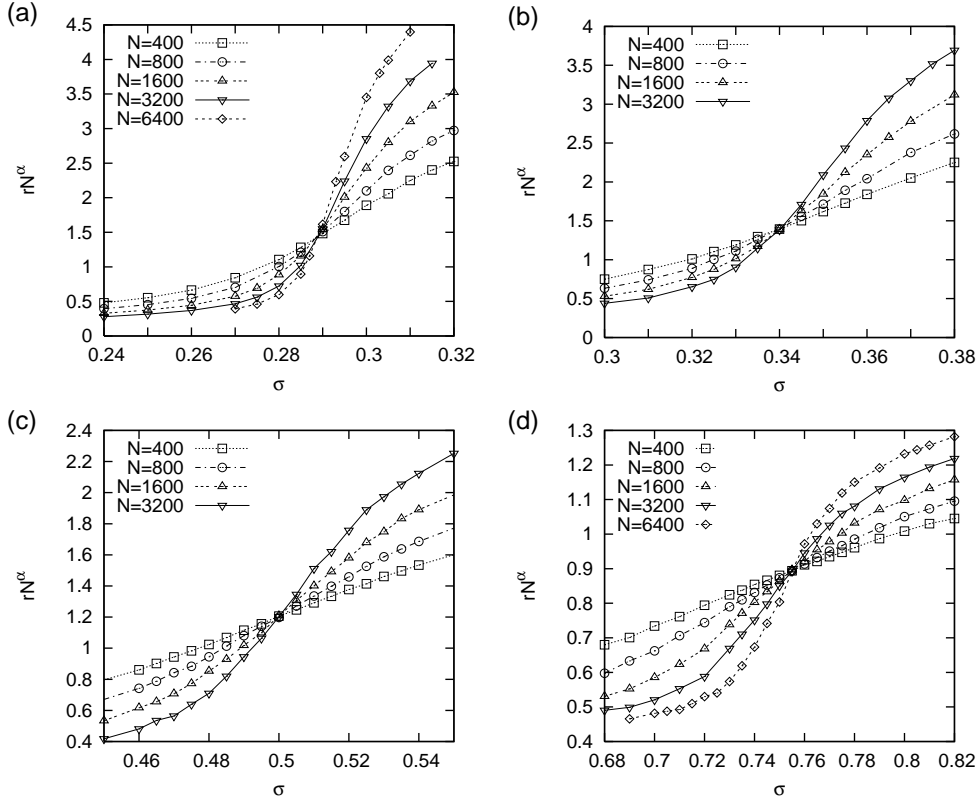}
 \caption{Determination of the critical points by FSS for $c_\omega=-0.6, -0.2, 0.2, 0.6$, panels (a)-(d) in the above order. The critical points obtained from the analysis are shown in Fig. \ref{figT}.}
% \caption{Determination of the critical points by FSS for $c_\omega=.6, .5, .4, .3, .2, .1,0,-.2,-.4,-.6$, panels (a)-(j) in the above order. The critical points obtained from the analysis are shown in Fig. \ref{figT}.}
\label{figF}
 \end{center}
 \end{figure*}

\begin{figure*}
 \begin{center}
\includegraphics[width=.95\textwidth]{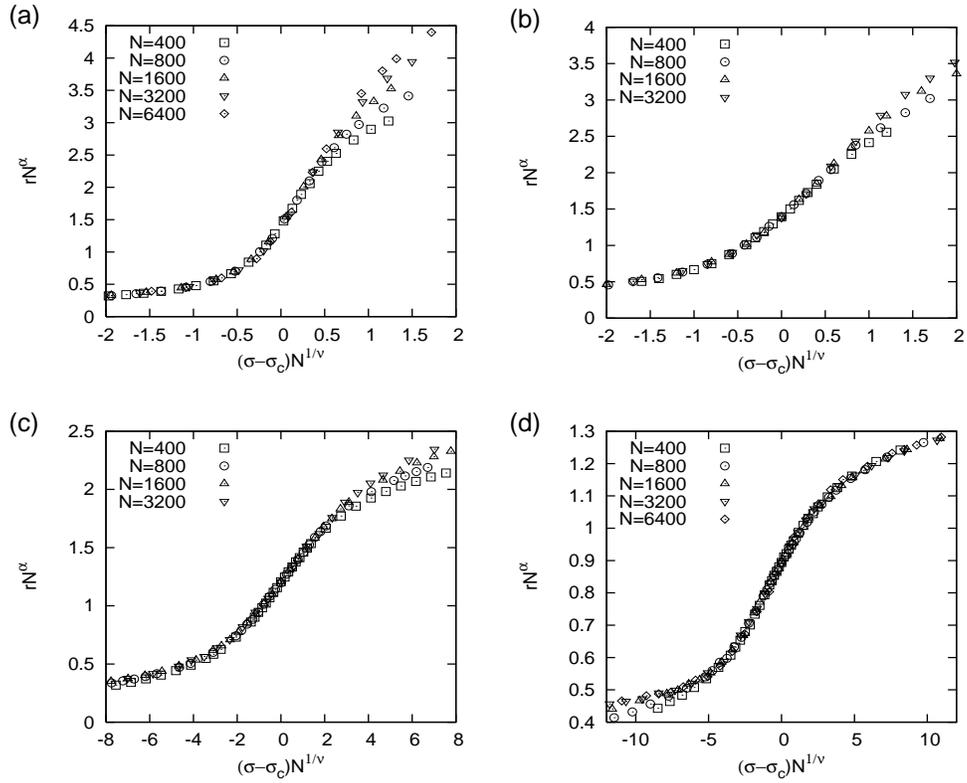}
 \caption{Data collapse close to the critical point for $c_\omega=-0.6, -0.2,0.2,0.6$, panels (a)-(d) in the above order. The critical exponents obtained from the analysis are shown in Fig. \ref{figT}.}
\label{figE}
 \end{center}
 \end{figure*}

Examples of this analysis are displayed in Figures \ref{figF} and \ref{figE}. Thus, for example for $c_\omega=0.6$ the curves $r(N)N^\alpha$ are found to intersect at a unique crossing point for $\alpha=0.075$ and a critical coupling of $\sigma_c=.755$ is obtained (top left panel of Fig. \ref{figF}). This result is confirmed by plots of $r(N)$ vs. $N$, which shows the algebraic dependence typical for a second order transition at this value of $\sigma$, cf. Fig. \ref{fig.1a}. For larger coupling the order parameter approaches a value independent of the system size, whereas $r(N)\sim {\cal O}(N^{-1/2})$ for smaller coupling. The analysis of the data collapse close to the transition point then yields $1/\nu= 0.6$, cf. top left panel of Fig. \ref{figE}. The remainder of the transitions for different oscillator correlations are analysed in the other panels of Fig's \ref{figF} and \ref{figE}.

\begin{figure*}
 \begin{center}
\includegraphics[width=.95\textwidth]{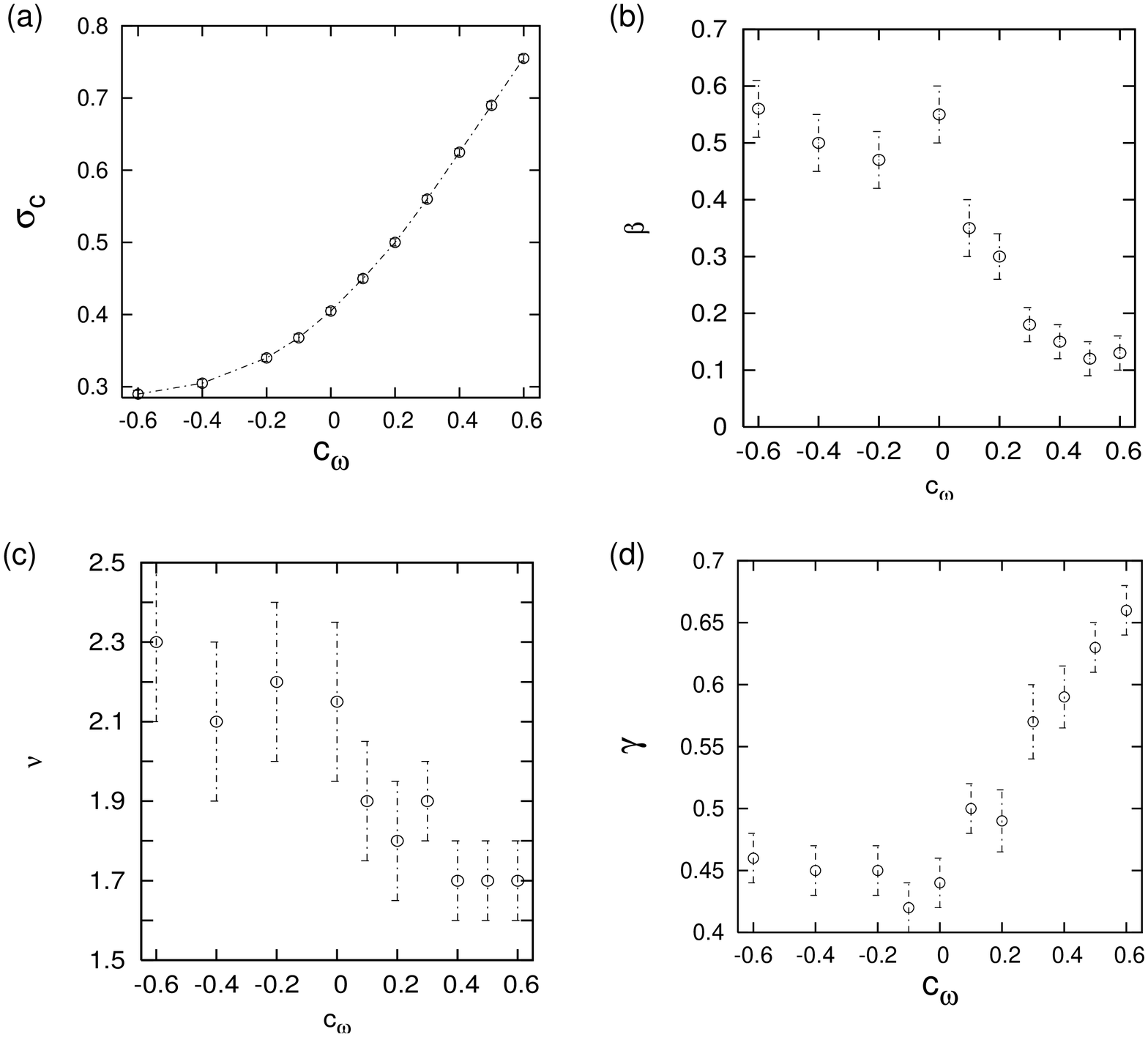}
 \caption{Dependence of the (a) critical coupling, (b) critical exponent $\beta$ and (c) critical exponent $\nu$ on the correlations between adjacent oscillators and (d) of the exponent $\gamma$ that characterizes the power law scaling of relaxation times with the system size in the critical region. Results are from the FSS analysis, lines are merely guides for the eye.}
\label{figT}
 \end{center}
 \end{figure*}

Following this FSS procedure for systematically varied values of oscillator correlations on the networks gives the phase diagram summarized in Fig. \ref{figT}a and the dependence of the critical exponents $\beta$ and $\nu$ on the oscillator correlations, cf. Fig. \ref{figT}b,c. The analysis clearly shows a monotonic dependence between oscillator correlations and critical coupling, i.e. the more anti-correlated the oscillator arrangement the less coupling is required for synchronization. 

Figure \ref{figT}a also indicates that possibilities to induce shifts to the onset of synchronization by oscillator anti-correlations are limited:  The critical coupling appears to saturate for $c_\omega\to -1$. Conversely, the growth of the critical coupling with $c_\omega>0$ is well-described by an exponential dependence $\sigma_c(c_\omega)=\sigma_c(0) e^{c_\omega/c_0}$ with $\sigma_0=0.96\pm.01$. 

One also notes a systematic dependence of the critical exponents $\beta$ and $\nu$ on oscillator correlations. For $c_\omega\leq 0$ our data are consistent with $\beta=0.5$ and it is evident that $\beta$ declines monotonically with $c_\omega$ in the positively correlated regime. In the case of the exponent $\nu$ the picture is less clear. Our data suggest an exponent $\nu$ in the range between $2.0\leq \nu \leq 2.5$ for $c_\omega\leq 0$. Exponents for $c_\omega>0$ are significantly smaller with an estimate of $c_\omega=1.7\pm 0.1$ for strongly positive correlations.

Hence, negatively correlated and positively correlated oscillator arrangements appear to be fundamentally different regimes of the systems' organization. For all $c_\omega\leq 0$, the data indicate that the synchronization transition falls in the same class as the all-to-all coupled Kuramoto system, i.e. $\nu=2.5$ and $\beta=0.5$ and no systematic dependence on the strength of oscillator anti-correlations is found. Conversely, in the regime $c_\omega>0$ the critical exponents vary systematically as the correlation strength is changed. Stronger positive correlations in the oscillator placement entail a monotonic decrease in $\beta$ and a change in $\nu$ from $\nu=2.2\pm0.2$ to $\nu\approx 1.7\pm 0.1$. Hence, for positively correlated oscillator placement finite size effects are found to grow in strength.

\subsection{Clusters of synchrony}
In this subsection we present a heuristic explanation for the observed dependence of the synchronization threshold on oscillator correlations. The explanation bases on the path towards synchronization on networks, similar to the findings of \cite{Arenas} for the evolution of clusters of synchrony on different types of networks or the observations of a dependence of the onset of synchronization on details of the structure of directed networks advanced in \cite{Me}. 

We define synchronized clusters in the following way. Two oscillators can be considered as synchronized when the difference in their time-averaged frequencies $\overline{\omega}=\lim_{T_\text{rel}\to \infty} 1/(T-T_\text{rel})\int_0^T \dot{\phi} dt$ is smaller than the frequency resolution of the numerical simulation, which is given by $1/T$. Accordingly, we define synchronized clusters as maximum sets of synchronized pairs of nodes. 

\begin{figure}
 \begin{center}
\includegraphics[width=.45\textwidth]{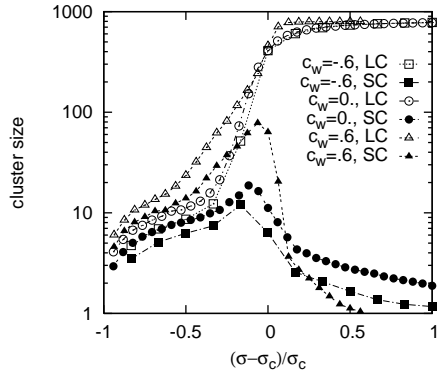}
\caption{Dependence of the average size of the largest (LC) and second largest (SC) synchronized cluster on the relative order parameter $(\sigma-\sigma_c)/\sigma_c$. The data are from experiments for a system of $N=800$ nodes and are averaged over 500 independent runs.}
\label{figCl}
 \end{center}
 \end{figure}

\begin{figure}
 \begin{center}
\includegraphics[width=.45\textwidth]{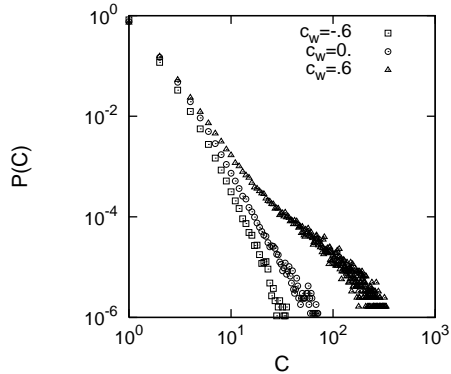}
\caption{Distribution of the sizes of synchronized clusters for $(\sigma-\sigma_c)/\sigma_c=-.05$. The data are from experiments for a system of $N=800$ nodes and are averaged over 500 independent runs.}
\label{figCl1}
 \end{center}
 \end{figure}

Figure \ref{figCl} shows simulation data for the dependence of the largest and the second largest synchronized cluster on the relative order parameter $(\sigma-\sigma_c)/\sigma_c$. It becomes apparent that while for positively correlated oscillator configurations large synchronized clusters are already formed far below the threshold, synchronized clusters are very small below the threshold for neutrally and even more so for negatively correlated configurations. Moreover, in strongly positively correlated ensembles, close to the transition several relatively large synchronized clusters not much smaller than the largest synchronized cluster exist, cf. Figures \ref{figCl} and \ref{figCl1}. This illustrates two different ways of cluster formation, similar to the differences in the formation of synchronized clusters on random graphs and scale-free networks described in \cite{Arenas}.

By definition, on positively correlated oscillator configurations, oscillators of similar native frequency tend to be neighbours. Hence, only a small coupling strength is required for them to synchronize. Thus, the regime below the transition to global synchrony is already marked by the presence of relatively large synchronized clusters. In each of these clusters oscillators support each other in a collective oscillation, which is mostly different from the global mean frequency $\Omega=1/N\sum_i \omega_i=0$.

In contrast, on negatively correlated configurations the formation of synchronized clusters below the transition point is suppressed. This is due to the large differences in native frequencies of adjacent oscillators, which makes it hard to form synchronized clusters for low coupling. Further, because of the negative correlations of native frequencies between adjacent oscillators, each of the small synchronized clusters that form nevertheless, has a collective frequency close to the global mean $\Omega=0$.

The transition to global synchrony requires the breakup of the local clusters to form one large synchronized cluster with a collective frequency equal to the global mean frequency. This, however, requires more coupling strength for positively correlated configurations because (i) the synchronized clusters that have to be merged are of roughly similar size and (ii) the collective frequency of each of these clusters tends to be substantially different from the global mean. 

Figure \ref{figCl} illustrates that this leads to two different paths to synchronization. In positively correlated configurations one synchronized cluster gradually merges with smaller synchronized clusters in its vicinity. In its essentials this mechanism is similar to the emergence of synchronization on scale-free networks \cite{Arenas}. On negatively correlated configurations the giant synchronized cluster emerges from the merger of many small synchronized clusters at the transition, similar to the path towards synchronization on random graphs.

\section{Relaxation dynamics}

\begin{figure}
 \begin{center}
\includegraphics[width=.45\textwidth]{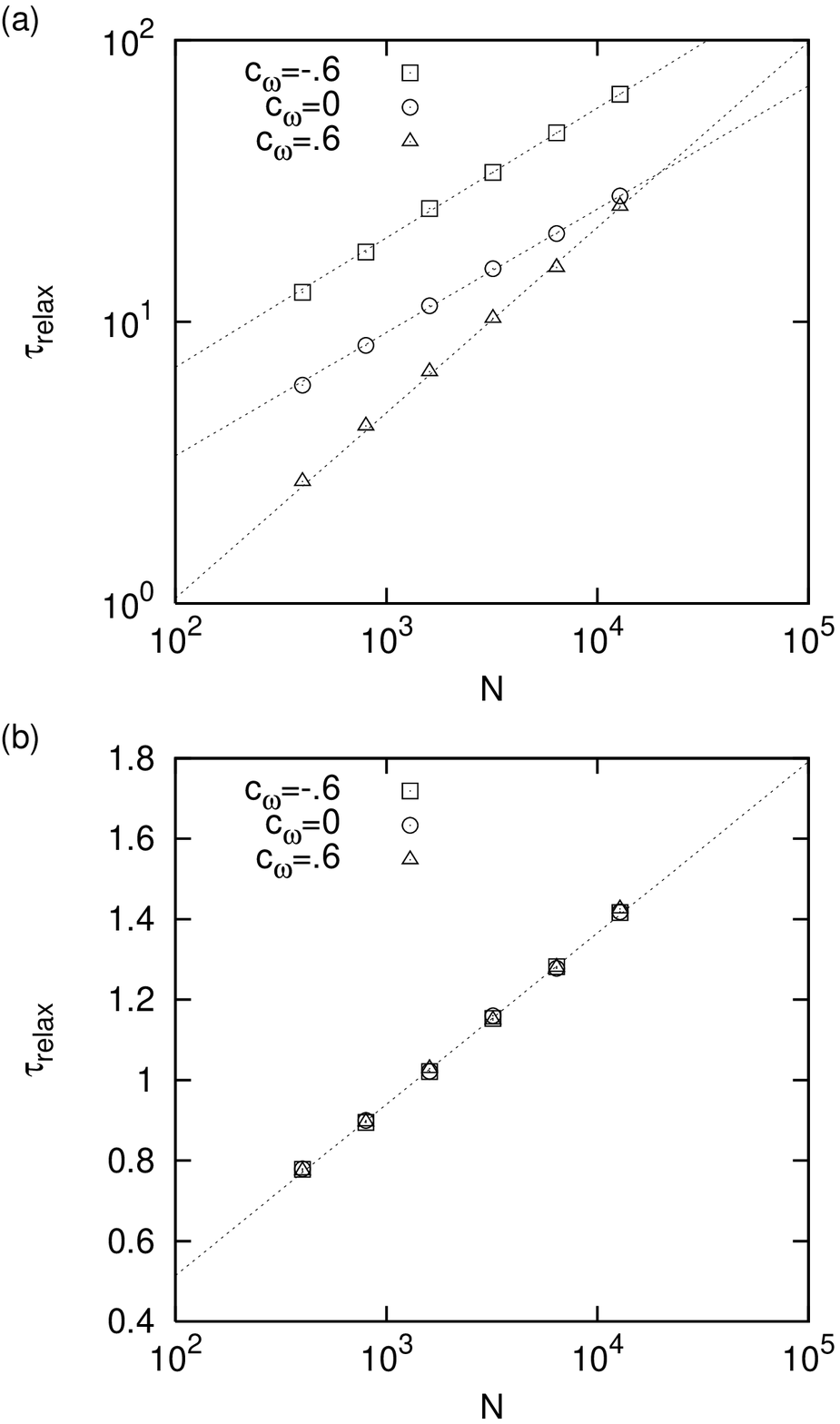}
 \caption{Scaling of relaxation times with system size for negatively (open squares), neutrally (open circles) and positively correlated(open triangles) oscillator arrangements.  (a) Near the critical points, power law behaviour is observed with $\gamma \approx 0.46$, $0.44$ and $0.66$. The behaviour for differently correlated oscillator arrangements is distinct. (b) in the strongly coupled regime ($\sigma=3.\gg \sigma_c$). The scaling of relaxation times with the system size is logarithmic, there is not discernable difference between relaxation times in differently correlated oscillator arrangements.}
\label{fig.Trelax}
 \end{center}
 \end{figure}

As we pointed out before, correlations of the oscillator arrangement on a network also influence the timescales of the relaxation dynamics towards the stationary state. This problem, the relaxation dynamics on networks in the strongly coupled regime, has recently found attention in \cite{Son}. Son et al. report that typical relaxation times scale logarithmically with the system size in the strongly coupled regime, but appear to follow a power law close to the critical point. Interestingly, this study also demonstrates that the logarithmic scaling is due to the heterogeneity in the initial phases and independenent of network characteristics.

For our analysis here, we follow the procedure introduced in \cite{Hong,Son} and renormalize the order parameter to
\begin{align}
 \tilde{r}=\frac{r_\infty-r(t)}{r_\infty-r_0},
\end{align}
where $r_\infty=\lim_{t\to\infty} r(t)$ and $r_0=r(0)$ and then calculate
\begin{align}
 \tau_\text{relax}=\int_0^\infty dt^\prime \tilde{r}(t^\prime).
\end{align} 
As pointed out in \cite{Son}, this definition of a relaxation time has the advantage of being robust for obtaining a saturation time and is also useful for a system with many relaxation time scales.

In Fig. \ref{fig.Trelax} we plot the scaling of the relaxation times with the system size for two different regimes, near the critical coupling strength $\sigma \approx \sigma_c$ and in the strongly coupled regime for $\sigma\gg \sigma_c$. Close the critical coupling, we observe the power law relation
\begin{align}
\tau_\text{relax} \propto N^\gamma
\end{align}
which was already mentioned in \cite{Son}. As can be seen from the top panel of Fig. \ref{fig.Trelax} the exponent $\gamma$ depends on the oscillator arrangement. For strongly negatively correlated arrangements ($c_\omega=-0.6$) a fit yields $\gamma=0.46\pm0.01$, for neutral correlations ($c_\omega=0$) a very similar exponent $\gamma=0.44\pm0.01$ is obtained, whereas for strongly positive correlations ($c_\omega=0.6$) with $\gamma=0.66\pm.02$ a clearly distinct behavior is recorded. Even though relaxation times are generally found to be much smaller for positively correlated oscillator arrangements in small systems, the large scaling exponent implies longer relaxation times in large systems. 

The exponents for oscillator arrangements with other correlations are displayed in Fig. \ref{figT}d. Even though error bars are relatively large the pattern appears to corroborate the previous observation of the existence of two distinct regimes of oscillator arrangements, $c_\omega>0$ and $c_\omega\leq 0$. For negative and neutral correlations the data indicate that $\gamma \approx 0.45$ independent of the correlation strength, whereas a monotonic growth of $\gamma$ with $c_\omega$ is manifested for positive correlation strength.

Whereas correlations in the oscillator arrangement thus clearly affect the behaviour in the critical region, relaxation times in the strongly coupled regime appear to be unaffected (bottom panel of Fig. \ref{fig.Trelax}). For $\sigma\gg \sigma_c$ in agreement with the results and derivation of \cite{Son} logarithmic scaling 
\begin{align}
\tau_\text{relax} \sim \tau_0 + \text{const.} \times \ln N
\end{align}
of the relaxation times with the system size is found. Data points for systems with differently correlated oscillator arrangements almost coincide, thus manifesting that the relaxation behaviour far above the critical point is independent of the oscillator correlations. 

\section{Discussion}

In this paper we have discussed the synchronization phase transition of systems of correlated non-identical Kuramoto oscillators on very sparse random networks. Negative correlations between adjacent oscillators are shown to decrease the critical coupling and allow for a transition to partial synchronization at lower coupling strengths than positively correlated oscillator arrangements. The detailed analysis of the critical couplings shows that the benefit from negatively correlated oscillator arrangements for the onset of synchronization saturates quickly with the strength of anti-correlations, whereas the increase of the critical coupling with the positive correlation strength is well characterized by an exponential dependence. These results confirm the previous findings of \cite{MB1} and demonstrate that the previous observations are not due to the relatively small system sizes, but are valid in the thermodynamic limit. As one of the key points of the present study this highlights that synchronization properties of a coupled system depend strongly on both, the coupling topology and the arrangment of the oscillators on the coupling network.

Importantly, the analysis of the phase transitions for positively and negatively correlated oscillator arrangements reveals two fundamentally different regimes: (i) for anti-correlated oscillator placements the transitions is still characterized by the critical exponents $\beta=1/2$ and $\nu=2.5$ as for the uncorrelated and the all-to-all coupled Kuramoto system, whereas for (ii) positively correlated oscillator placements the critical exponents are found to vary systematically with the correlation strength $c_\omega$. The effect of positively correlated oscillator arrangements is thus comparable to a change in the heterogeneity of the coupling network.

We also investigated the relaxation behaviour towards the stationary states of the dynamics. Close to the critical point, where power law scaling of the relaxation times with the system size $\tau_\text{relax} \propto N^\gamma$ is found, oscillator correlations strongly influence the relaxation dynamics. For negative and neutrally correlated oscillator ensembles we find $\gamma\approx 0.45$, whereas for strongly positively correlated ensembles the best estimate is $\gamma \approx 0.66\pm0.02$. Conversely, far above the critical coupling strength, the scaling of relaxation times with the system size is logarithmic and no dependence on correlations in the oscillator arrangement is found.

The results presented in this paper raise a number of interesting questions about the influence of oscillator correlations on the synchronization transition for other coupling arrangements. For instance, it would be of interest to investigate the influence of oscillator correlations on the critical behaviour on scale-free networks, a topic of quite some interest over the last years \cite{Moreno,Lee,Hong2,Gardenez}. By the same token for the uncorrelated case it has been established that the lower critical dimension for phase synchronization on hypercubic lattices is $d^P_l=4$, whereas it is $d_l^F=2$ for frequency entrainment \cite{Hong1}. Could oscillator correlations influence the lower critical dimensions for phase and frequency synchronization on hypercubic lattices? We think that both issues are interesting directions for future work.


\begin{thebibliography}{0}
\bibitem{remark}	
         Our numerical investigations show that fixing the number of links significantly reduces integration times and ensemble sizes required to reduce fluctuations.
\bibitem{Bem0}
         We also experimented with a Gaussian distribution for the native frequencies. Even though we do not present detailed simulations for this case here, the results presented below were found to be robust and not specific to the choice of frequency distribution. For instance, for networks with $\langle k\rangle=10$ and native frequencies distributed according to a Gaussian distribution centred around zero with variance $0.5$ we found $\sigma_c=0.084(1)$, $\beta\approx 1/2$ and $\nu\approx2$ for $c_w=0$, whereas $\sigma_c=0.166(1)$, $\beta\approx 0.08$, $\nu \approx 1.65$ for $c_w=0.5$ and $\sigma_c=0.064(1)$, $\beta\approx 1/2$ and $\nu\approx 2$ for $c_w=-0.5$.
\end{thebibliography}
\end{document}